\documentclass[traditabstract]{aa}
\usepackage{graphicx}

\usepackage{txfonts}

 \usepackage{natbib}
 \bibpunct{(}{)}{;}{a}{}{,} 

\newcommand{\ephi} {{\bm {\hat e}}_\phi} 

\newcommand{\dd}{\mathrm d}

\newcommand{\kep}{{{\it Kepler }}}
\newcommand{\kic}{{{KIC~5955122~}}}
\newcommand{\teff}{{{T_{\rm eff}}}}

\def\rot{\nabla\times }

\def\eT{\eta_{t}}

\def\be{\begin{equation}}
\def\ee{\end{equation}}    
\def\ba{\begin{eqnarray}}
\def\ea{\end{eqnarray}}    

\def\bm{\mathbf}

\usepackage{color}

\newcommand{\ins}[1]{\textcolor{blue}{#1}}
\usepackage{epstopdf}

\begin{document}

\title{Magnetic activity, differential rotation and dynamo action in the pulsating F9IV star \kic}
\author{A. Bonanno\inst{1}\and
H.-E. Fr\"ohlich\inst{2}\and
C. Karoff \inst{3,4}\and
M.~N.~Lund\inst{3}\and
E. Corsaro \inst{5,1}\and
A.  Frasca \inst{1}
          }
\offprints{Alfio Bonanno\\ \email{alfio.bonanno@inaf.it}}

\institute{
INAF, Osservatorio Astrofisico di Catania, via S. Sofia, 78, 95123 Catania, Italy
\and
Leibniz Institute for Astrophysics Potsdam (AIP), An der Sternwarte 16, 14482 Potsdam, Germany
\and Stellar Astrophysics Centre, Department of Physics and Astronomy, Aarhus University, Ny Munkegade 120, DK-8000 Aarhus C, Denmark
\and Department of Geoscience, Aarhus University, H{\o}egh-Guldbergs Gade 2, 8000 Aarhus C, Denmark
\and
Instituut voor Sterrenkunde, KU Leuven, Celestijnenlaan 200D, B-3001 Leuven, Belgium\\
}

\date{}
%
%
\abstract{
We present photometric spot 
modeling of the nearly four-year long light-curve of the \kep target  \kic
in terms of persisting dark circular surface features. 
With a Bayesian technique, we produced a plausible surface map that shows dozens of small spots. After some artifacts are removed, the residuals are at $\pm 0.16$\,mmag.
The shortest rotational period found is $P = 16.4 \pm 0.2$ days. 
The equator-to-pole extrapolated differential rotation is $0.25 \pm 0.02$ rad/d. 
The spots are roughly half as bright as the unperturbed stellar photosphere. 
Spot latitudes are restricted to the zone $\pm 60^\circ$ latitude. There is no indication for any near-pole spots.
In addition, the p-mode pulsations enabled us to  determine the evolutionary status of the star, the extension
of the convective zone, and its radius and mass.  
We discuss the possibility that the clear signature of active regions in the light curve of the F9IV star
\kic is produced by a flux-transport dynamo action at the base of the convection zone. 
In particular, we argue that this star has evolved from an active to a quiet status during the Q0--Q16 period of observation, and 
we predict, according to our dynamo model, that the characteristic activity cycle is of the order of the solar one.}
\titlerunning{Magnetic activity and differential rotation in \kic}
      \authorrunning{Bonanno}

\maketitle
%
\section{Introduction}
The high-quality photometric data from the {\it CoRoT} \citep{baglin06} 
and  {\it Kepler} \citep{borucki10} space telescopes have opened new possibilities for investigating 
the origin and evolution of magnetic fields in solar-type stars.

In particular, the possibility of obtaining almost uninterrupted observations for periods  much longer than the typical rotation periods
has proven to be a key ingredient for characterizing the photometric inhomogeneities and their evolution in young, 
fast-rotating objects \citep{fro09,bola,frasca2011,froehlich2012}. 
In solar-like stars the frequency shifts and amplitude modulations in the 
p-mode spectrum represent an important diagnostic for detecting magnetic activity cycles, 
as was recently shown for the F5V star HD 49933  \citep{garsci} and in the subgiant $\beta$ Hyi  \citep{trew}.
In these cases a theoretical interpretation of the p-mode
spectrum can determine the  precise evolutionary status of the star and, in some cases, its internal distribution of angular momentum
and the inclination axis as well.

From the theoretical side, it is generally expected that stars with deep convective envelopes (late F, G, K, and
early M spectral type) produce a strong radial shear at the interface between the convective zone and the radiative interior,
and in fact, numerical simulations of young solar-type stars have clearly demonstrated the possibility of an efficient dynamo action in this case \citep{aug}.
The situation is less clear for  early F-type stars where the shallow convection zone 
makes it unlikely that a large-scale azimuthal field is stored near the surface \citep{parker93}.
On the other hand, it has been shown \citep{huber} that the subgiant F5IV-V star Procyon A is significantly more  active than the Sun, although its convection zone is only 
$\approx 3-4\%$ in radius.
A recent work based on a sample of 22 solar-type F stars observed by \kep has also shown that the activity level in evolved F-stars is in general stronger than expected \citep{mathur14}. The key question is therefore which type of dynamo action is at work in this class of objects.

To address this question we  have taken advantage of the exquisite quality of \kep photometry for KIC~5955122,  a solar-like F-type pulsator with a clear signature of spot 
rotational modulation and spot evolution.  The key idea is to explicitly build a dynamo model of the magnetic field in this object in the framework 
of a mean-field approach, and show that its predictions can be compared with the solution of the spot modeling for the photospheric field.

Although this task is in general very difficult, the detection of an acoustic oscillation spectrum in \kic is a powerful diagnostic for 
the stellar interior, and in particular, for the 
convection efficiency.  According to \cite{appo12}, this object has a p-mode 
spectrum that is characterized 
by a large separation $\Delta\nu = 49.6 \; \mu {\rm Hz}$, with an excess of power around $\nu_{\rm max} \sim 826 \; \mu {\rm Hz}$,
which enables estimating the mass, radius, and convective velocities.
Because the information on the rotation rate and surface differential rotation is obtained from the spot modeling,
reasonable  estimates of all the characteristic dynamo numbers are available in this case.

Today, the most promising theoretical framework  within which the dynamo action in the Sun is described is the  advection-dominated (or flux-transport) dynamo  \citep{dik09,guerrero,boso}.
According to this idea, the meridional circulation acts as a conveyor belt that  
transports the magnetic flux toward equatorial latitudes, provided the eddy diffusivity is low enough \citep{bo02}. 
It is then interesting to test this idea on our object because so far its application to a large class of solar-type Mt.Wilson active stars was shown to be problematic \citep{jouve10}.

We argue that a flux-transport dynamo can explain several features of the topology and the evolution of the photospheric  field in KIC~5955122.

The structure of the paper is as follows:  {Sect.~1} contains the introduction, in {Sect.~2} 
we describe the observations and data reduction, {Sect.~3} describes the Bayesian spot modeling,
and Sect.~4  contains the description of the  flux-transport dynamo model and a discussion of the numerical results. 
{Section~5} is devoted to the conclusions. 

\section{Observations of \kic}
\subsection{Ground-based data}
\kic (=TYC 3142-1229-1) has a \kep magnitude of $V=9.281$ ,  $\teff=5747$ {K,} and {$\log g = 4.3$ dex} 
according to the Kepler Input Catalog (KIC, \cite{kic}).
Because it has been shown that the values of $\teff$ given in
the KIC are systematically too low for solar-type stars \citep{pin12}, 
we used spectroscopic data 
obtained with ESPADONS on April 2010 to derive the stellar astrophysical parameters. 
We used the code ROTFIT \citep{frasca03,frasca06}
to evaluate $\teff$, $\log g$, [Fe/H], and determine $v \sin i$. ROTFIT uses a standard $\chi^2$ minimization on individual orders 
using a library of 185 
ELODIE archive spectra of standard stars, which covers the space of atmospheric parameters for FGK-type stars with 
a metallicity above -1 \citep{prug01} very uniformly. The result is $\teff=5954 \pm 70 \; {\rm K}$, 
{$\log g = 4.13 \pm 0.18\; {\rm dex}$}, [Fe/H]$=-0.06\pm 0.12 \;{\rm dex}$\ins{,} and 
$v \sin i =4.4 \pm 1.5 \; {\rm km \;s}^{-1}$. The spectral type is F9IV, which indicates that this is an evolved subgiant. 

Another spectroscopic study \citep{bruntt12} instead reported $\teff=5865 \pm 60 \; {\rm K}$ 
and {$\log g = 3.88 \pm 0.2\; {\rm dex}$}, {and} $v \sin i =6.5 \; {\rm km \;s}^{-1}$ in agreement with 
the values provided by ROTFIT within the {uncertainties}.
The differences between the values obtained in \cite{bruntt12} and the values adopted in this work result in slightly different 
limb-darkening coefficients, which does not affect the results of the spot modeling in any way, however. 

To obtain an asteroseismic estimate of the stellar mass and radius we employed  the code Catania-GARSTEC \citep{bosh02}, 
following the approach and the physical input described in  \cite{met10} for  KIC~11026764.
In our grid model search  we considered the average values  from ROTFIT and \cite{bruntt12} as astrophysical constraints, 
and the large separation $\Delta\nu = 49.6 \; \mu {\rm Hz}$  \citep{appo12} as the main asteroseismic constraint.

The resulting best model has $M=1.12 M_\odot$, $\teff = 5904 \;{\rm K}$, $\log g  = 3.87$ dex , $\Delta\nu = 49.63 \; \mu {\rm Hz}$, [Fe/H]$=-0.11$, 
$R=2.0 R_\odot$\ins{,} and an age of $5.223$ Gyr{. This is} 
very similar to the solution obtained in \cite{met14} using the values of \cite{bruntt12} as astrophysical constraints and fitting the individual frequencies. 
The extension of the convective zone is also very similar to that of the Sun, being $r_{\rm cz}=0.75 R_{\ast}$. 
It is thus reasonable to think  that in this star a tachocline 
is also present at the interface between the radiative and convective zones. 

\begin{figure}[htp]
  \includegraphics[width=8cm]{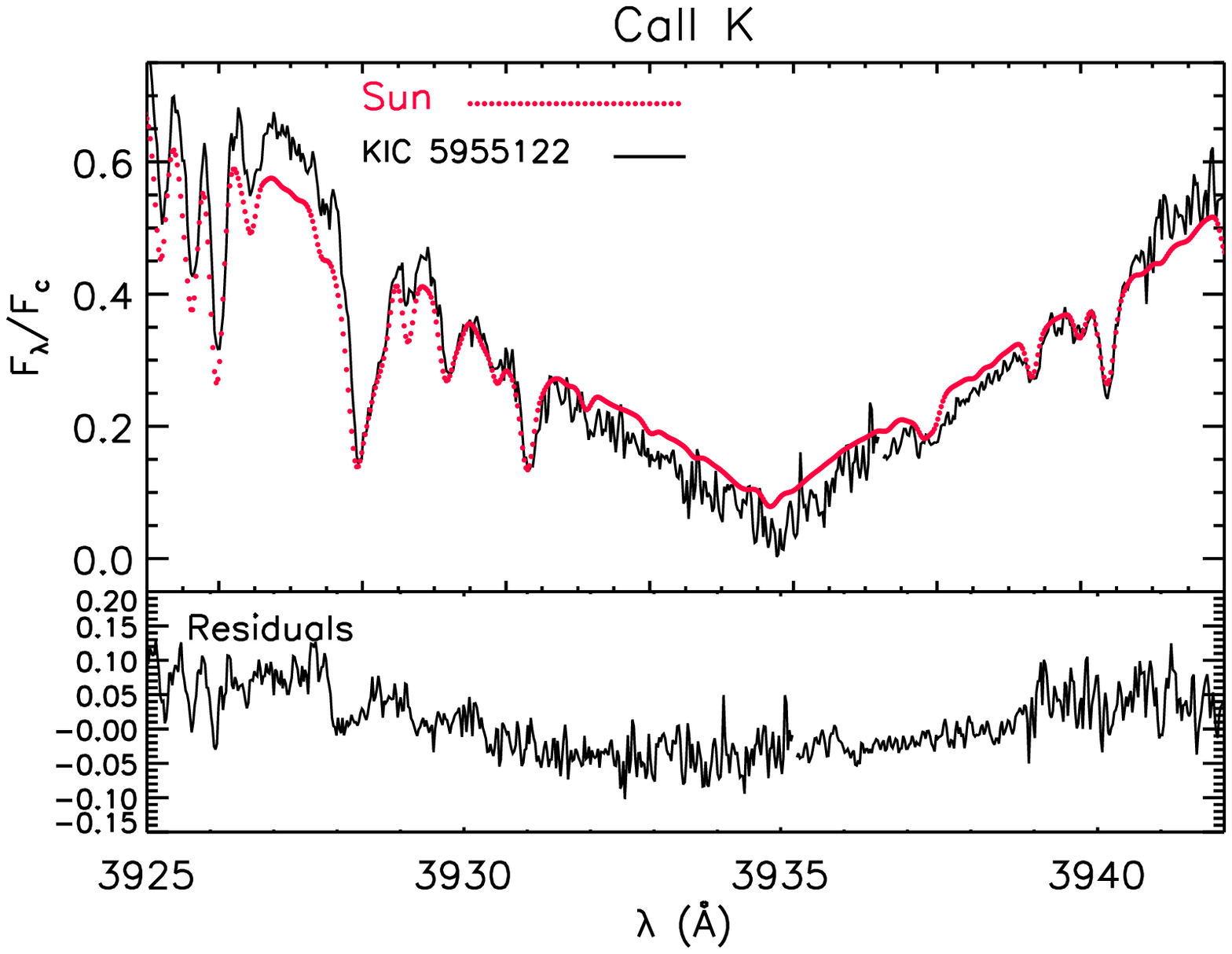}   
   \include graphics[width=8cm]{cah.eps}   
  \caption{Continuum-normalized observed spectrum (solid black line) 
 in the region between 3925 and 3942 \AA ~({\it top panels})
and between 3960 and 3975  \AA ~({\it bottom panels}) of KIC~5955122, 
compared with the Sun near the minimum of the activity cycle (HARPS spectrum of Ganymede taken in 2007).
\label{f1}}
\end{figure}

Additional observations have been carried out on 26 August 2013 with the high-resolution FIbre-fed Echelle Spectrograph (FIES) mounted on the 2.6-m Nordic 
Optical Telescope \citep{2000mons.proc..163F, 2014AN....335...41T} to measure the chromospheric activity level. We obtained five spectra with seven-minute exposures using the low-resolution {fiber} (R=25,000), which resulted in a S/N of about 100 at the blue end of the spectrum. Data reduction was performed with FIEStool\footnote{http://www.not.iac.es/instruments/?es/?estool/FIEStool.html} following the basic steps described in \cite{kar13}. The values obtained for the Mt. Wilson $S$-index in the five spectra 
{are}
$S=0.146, 0.181, 0.122, 0.148,      0.142$. If we exclude the value of $0.181,$ which seems to be caused by a cosmic-ray hit in the spectrum,
we finally obtain $S=0.140\pm 0.007$. Compared with the Sun, the chromospheric emission is similar to that of a quiet Sun, as is apparent from Fig.~\ref{f1}.

\subsection{\kep data}
The photometric data used in this work were collected by the \kep photometer in the 
period from May 2009 to April 2013, corresponding to the run 
Q0--Q16 for long-cadence data. {The light curve was constructed and corrected for according to the method described 
in Handberg~\&~Lund (in prep.). Briefly, flux was extracted from \textup{\emph{target pixel files}} and bad}  data points {were} 
removed according to the flags from the \kep team. {Only a mild de-trending was applied ($\tau_{\rm long} = 50$ days and $\tau_{\rm short} = 10$ days), with the main effect being the removal of the yearly modulation from the roll of the spacecraft. We refer to Handberg~\&~Lund (in prep.) for more details on the methodology used for correcting the light curve.} 

\begin{figure}[htp]
\begin{center}
  \includegraphics[width=7cm]{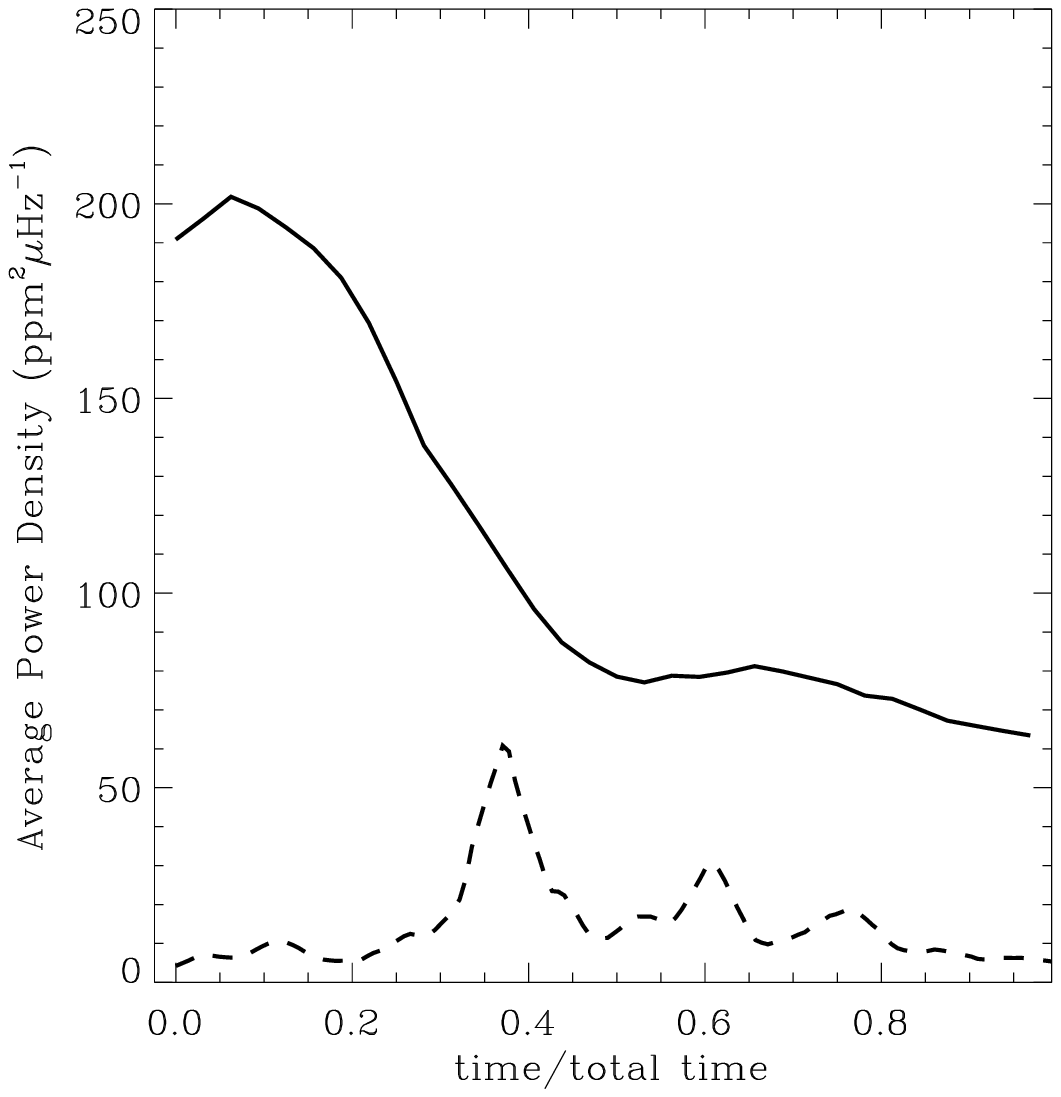}   
   \includegraphics[width=7cm]{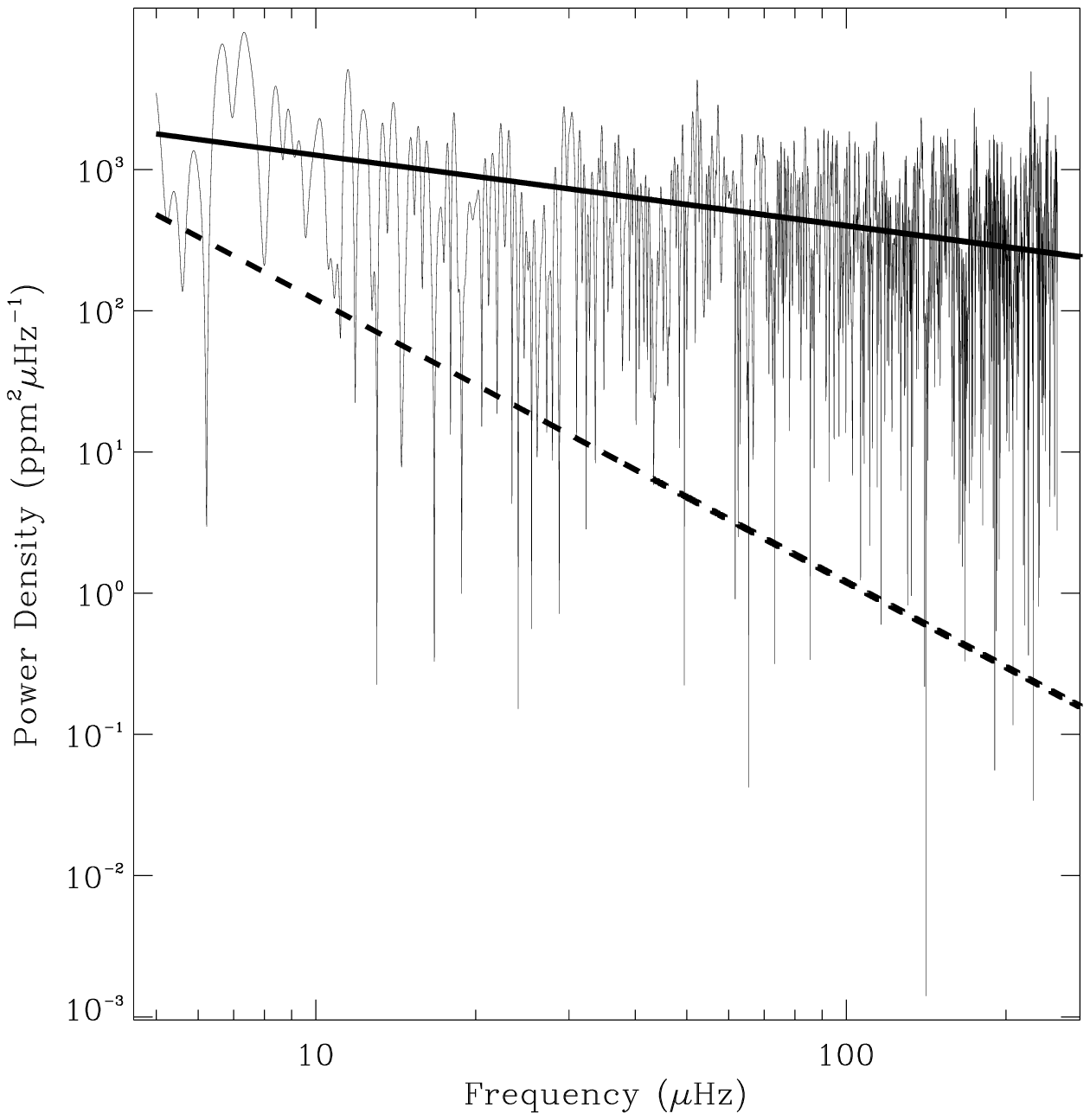}   
   \end{center}
\caption{Upper panel: average power density as a function of time in units of the total time for 
\kic (solid line) and for the Sun (dashed line). A prominent peak around 0.36 (in units of the total time) 
{is seen for the Sun and corresponds}  to the maximum of solar cycle 23.
Lower panel: power density at the maximum of the upper figure for \kic (continuous line) and for the Sun (dashed line, fit to the power density). Note that  in {Eq.~}(\ref{qqq}) 
$\beta=1/2$ for \kic and $\beta=2$ for the Sun.
\label{f2}}
\end{figure}

It is interesting to study the time evolution of the power spectral density in the low-frequency region. 
We  divided  the full observational run into chunks of 20 days (a convenient time interval about the same length as the rotational period, as can be easily seen by direct inspection of the light curve)
and computed the photometric power spectral density ${\rm PSD}(\nu)$ in each interval.
As argued in \cite{huber}, this quantity is proportional to the fractional area covered by active regions, $d a / a$,
and to the luminosity variations $\delta L/L$  caused by  the flux contrast between the unspotted and spotted stellar areas \citep{dorren}, 
\begin{equation}
{\rm PSD}(\nu) = \left ( \frac{da}{a} \right)^2 \left ( \frac{\delta L/L}{T}\right ) \nu^{-\beta}
\label{qqq}
,\end{equation}
where $\beta \approx 2$ for the Sun.
The result is shown in Fig.~\ref{f2}, in which we have plotted  the average PSD in the region 5--250 
$\mu$Hz in the top panel as a function of time in units of the 
total time of run Q0--Q16 (about four years).  During the first part of the run (mostly during runs Q1 and Q2) 
there is a clear excess of power that gradually decreases at later times. For comparison,  the dashed line depicts 
the same quantity for the Sun for the VIRGO observations in the 
green channel \citep{frov}, obtained with the {\it SOHO} spacecraft, for the period 
1996--2004. 
In this case, the excess of power is now always significantly lower 
than for KIC~5955122, 
even during the activity maximum of cycle 23, corresponding of about 0.36 in the normalized {time} units of the plot.
We argue that if this star has
an activity cylce, the evolution of the PSD can be explained as a transition from a maximum of the activity cycle
to a more quiet status (during the Q10--Q16 period) where the corresponding chromospheric activity $S$-index is rather low,
as expected. 

On the other hand, the star is significantly more active during its activity maximum than the Sun. 
We compare the PSD in the 5--250 $\mu$Hz {region} during the maximum in 
Fig.~\ref{f2} (upper panel) with the same quantity for the Sun taken at maximum of cycle 23. 
The result is shown in {the lower panel of} Fig.~\ref{f2}  
where the dashed line, obtained by fitting the PSD decay for the Sun, is to be compared with the solid line obtained for KIC~5955122.
In particular, $\beta\approx 2$ in Eq.~(\ref{qqq}) for the Sun, while $\beta\approx 0.5$ for KIC~5955122.

\section{Bayesian spot modeling}
\begin{figure*}
\centering
\includegraphics[width=17cm]{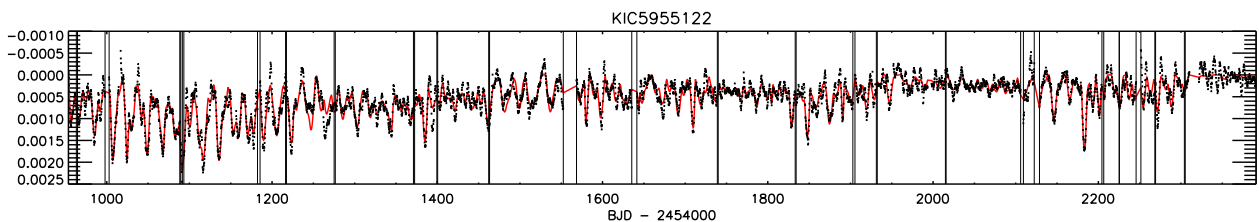}
\vspace*{-9cm}
\caption{ 5335 data points from Q0 to Q16 fitted by a 33-spot model (red line). 
The light curve was divided into 25 parts, whose boundaries are indicated by vertical lines. 
The residuals are $\pm 0.16$~mmag,  i.e., $87.5\%$ of the photometric variance is attributable to dark spots. 
}
\label{fig_lc}
\end{figure*}

{Equation~(\ref{qqq})} only provides a rough estimate of the excess of power caused by active regions{. To}  
characterize the evolution of the photospheric field we discuss
in this section a detailed spot model that assumes comparatively long-lived dark features. 

\subsection{Some general remarks}

For actual computations the nearly 60,000 long-cadence data points were combined into bins of 0.25 days width. 
Each of the 5335 final data points was assigned a weight according to the number of contributing original data points within a bin. 
Moreover, because the original data show gaps in time in addition
to the gaps caused by the measuring quarters Q0 to Q16, 
the whole data set was accordingly divided into 25 parts (see
Fig.3). 

In principle, the spot-modeling procedure 
is as described in \cite{frasca2011} and \cite{froehlich2012}. 
Central to this parameter estimation is the likelihood function. 
It measures the probability of the data given a set of parameter values. 
Because we{} split{} the light-curve into parts, the combined likelihood function 
is the product of {the} 25 contributions. For each  chunk {of the light curve} a 
partial likelihood function was constructed as follows:
{the} residuals, {given by} the deviations between observation{s} and {the} 
theoretical model, with measurement errors augmented by 
systematic (model) errors, were {assumed} to be 
Gaussian-distributed in magnitude with unknown 
variance. Because the data points (bins) were assigned different weights, 
a data point's variance scales as the ratio of 
an unknown variance divided by the number
of original data points within the bin. 
 To exclude this unknown variance, 
the Gaussian likelihood function was integrated over all conceivable values of this unknown variance 
by applying Jeffreys' prior, meaning that 
the unknown variance's prior behaves 
like the reciprocal of the variance itself. 
Afterward, this error-integrated likelihood was additionally integrated over all conceivable magnitude off-sets so that  it was not necessary  to specify any zero point.
Both integrations were made analytically \citep{fro09}. 
It is important to note that each part has its own characteristic 
error variance and  magnitude off-set. Because both quantities are regarded as mere 
nuisance parameters, they remained undetermined. 
Only afterward, for presentational purposes, when the data are overplotted on the theoretical 
light-curve (Fig.~\ref{fig_lc}), the off-sets
need to be fixed. This was made
by minimizing the distance between the data and the theoretical light-curve. 

Theoretical light-curves were computed using   
the analytical formulae from \citet{dorren} for circular spots, but they were generalized for quadratic limb-darkening (LD)

The two coefficients of the quadratic  LD relation 
were {adopted from \citet{Claret}}
assuming an effective 
temperature of $T_{\rm eff} = 5954$\,K, a gravity of 
$\log g = 4.13$ dex, and solar metallicity. For the LD 
no difference was made between the unperturbed photosphere and spots. 

Our aim is to analyze the photometric data in terms of a model 
with lasting dark spots that differ in rotational period to derive
at least 
a lower limit of the amount of differential rotation 
needed to fit the photometric measurements by a spot model. 
The spots were allowed to evolve. 
To estimate the differential rotation we assumed that
the spots survive at least a few rotations.
Hence, for the sake of simplicity we tried to explain 
the measured light curve with as low a number of spots 
as feasible, {that is,} ideally with the smallest number of free parameters 
necessary to reach a given fit accuracy. The Bayesian goodness-of-fit by varying the number 
of free parameters may be quantified by applying the 
{\citep[Bayesian information criterion  (BIC;][]{Schwarz}.} 

\subsection{Search for photometric trains of dips}
\begin{figure}
\includegraphics[width=8cm]{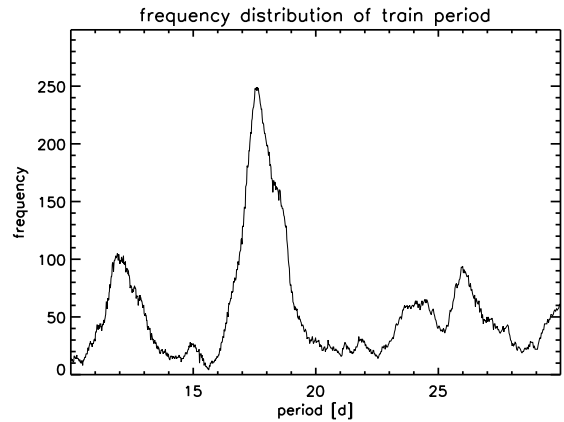}
\caption{Frequency distribution of train periods. We considered
only trains with at least five consecutive dips 
whose depths exceed 0.5 mmag.}
\label{fig_trains}
\end{figure}

The light curve (Fig.~\ref{fig_lc}) shows multiple spots. 
We searched for photometric trains of at least five consecutive
dips separated by equal time intervals. The darkening had to
exceed 0.5 mmag to be considered a dip in the light curve. This threshold was the only parameter in our train search. The period distribution for
trains{} shown in Fig.~\ref{fig_trains} for periods between 10 and 30 days seems to be bi-modal.
Whether a certain train is due to 
the cyclic appearance of a physical spot or occurred by chance cannot be decided. 
We can only conclude from the period distribution
the absence of certain spot periods. 
If there are long-lived spots, at least some trains of spots must be 
real, that is, be{\it } caused by spots that come into view periodically, otherwise 
the train period distribution would have been much more erratic, as can be proven
by a Monte Carlo simulation.

A complication arises because the off-sets of different 
parts of the light-curve differ. As mentioned before, these off-sets 
can be only estimated in retrospect. Hence, the period distribution 
of the trains itself depends somewhat on the anticipated result of the spot modeling. 

Periods are most frequently found around 18 and 26 days. The short-period peak (near 12 days) is 
the second harmonic of the longer-period peak. 
From the width of the main peak,  which covers periods from 16 to 20 days,  we derived a lowest value 
of the differential rotation of 0.08 rad/d. 
The question arises whether spot periods of the secondary peak around 
26 days must be incorporated to derive an acceptable fit for the light curve or not, which results in a much longer 
differential rotation. 

\subsection{Details of the spot-modeling procedure}
The spot modeling started with the five longest trains with periods around 
18 days. While in the search for trains of regularly appearing dips the 
area evolution of the prospected spots can be arbitrary because
the darkening just needs to exceed a certain threshold, the area evolution  
in the subsequent spot-modeling procedure is specified. 
Basically, the waxing and waning of spot areas is assumed to evolve linearly with time. Hence, 
at least four parameters are needed: the largest spot area (expressed in units of the 
stellar cross-section), the time of largest extent, and two slopes, differing 
in sign, which describe the increase of the spot area before the time of largest extent 
and the subsequent decrease. 
To mitigate the sharp bend at the time when the spot occupies
the largest area, another parameter 
is introduced: the length of a time interval. Within this time interval 
the actual derivative of the area vs. time relation is linearly interpolated 
between the two slope values adjacent to this time interval. 


{As explained in \cite{frasca2011}, it is convenient to use 
logarithmic quantities to compute posterior probabilites.}

A spot central longitude is given with respect to a rigidly rotating coordinate system 
with an arbitrary period of 18.5 days. {We define the longitude as increasing} in the direction of stellar
rotation{, with the zero-}point {given as} the central meridian facing the observer
at the beginning of the time series. Two free parameters 
need to be estimated to describe position and period: 
the spot longitude at the beginning of the light curve and 
at the end. The period follows from these two longitude values given the 
reference period. 

The (absolute) latitude of a spot center 
is computed from the period 
and the two parameters that specify the law of differential rotation. 
A simple $\sin^2$-ansatz was used (cf. \cite{frasca2011}, Eq.~2). 
The two parameters used are $\sin^2(\beta_{\rm s})$ and $\sin^2(\beta_{\rm f})$, 
where $\beta_{\rm s}$ and $\beta_{\rm f}$ describe the 
latitude of the fastest and of the slowest spot. 
{ In principle, our ansatz even includes antisolar rotation 
in line with the surface rotation profile $\Omega_{\rm s}(\theta)$, 
with $\theta = \pi/2-\beta$, which is exploited in Eq.~(\ref{eq3}) of Sect.~\ref{basics}. 
Here we only considered the solar-like case. 
Because we are interested in estimating
a lower bound on the differential rotation, the fastest spot 
was fixed to the equator. This reduces the number of free parameters and
therefore increases the credibility of the model.} 
The hemisphere {to which} a spot belongs  has to be {asserted} 
by trial-and-error.

Additional free parameters are the cosine of the stellar inclination, 
$\cos(i)$, and the common rest intensity, $\kappa$, of the spots, 
which is expressed in units of the surface brightness of the (spotless) star.

Each spot is described by a total of six free parameters. 
The seventh  parameter, which describes the smoothing 
of the otherwise sharp bend at the time of the largest extent
of a spot, 
was fixed by assuming a smoothing time span of 2.7 days.  
The hemisphere {of the spot} may be considered as an eighth parameter. 

The likelihood function 
in the high-dimensional parameter space already matches the posterior 
probability distribution because the parameters are already represented such 
that their prior distribution is a flat one in parameter space.

\subsection{Enlarging the number of spots}

Because it proved impossible to fit the light curve by the effect of 
a few enduring spots, we considered many spots 
with lifetimes considerably shorter than the time span of the 
measurements, but exceeding at least a few rotations. 
{We successively increased the number of spots, and for each added spot ran the adopted Markov chain Monte Carlo (MCMC) procedure anew}. 
A new spot was inserted to match
{a feature in the light curve that the spot-model so far did not account for.} 
Its initial period was chosen such that adjacent features 
were accounted for by the new spot {as well}. 
We always {started} with a {guess on the} period within 
the interval {from} 16 to 20 days. Only when this procedure prove{d} to be impossible did we assume  a 
longer initial period. 
By adding {an additional} spot, all the other spots that contribute 
to a periodic darkening in that time span of the light curve 
have to adjust to the new setting. This is a highly nonlinear 
problem because many dips are caused by the cumulative effect of more 
than {a} single spot. 
In the end,  we obtained a {\it \textup{possible}} solution.
This proves that it is at least feasible 
to interpret the data to some degree within the framework of our 
model assumptions. 
There may be a plethora of possible 
solutions, however. In the worst case, all spots are ephemeral,  and 
{interpretating} photometric trains in terms of 
periodically appearing and disappearing spots is  misleading. 
{However}, with solely ephemeral spots 
the smoothness of the bi-modal frequency distribution of photometric trains 
of different periods (Fig.~\ref{fig_trains}) is hardly conceivable. 
It seems reasonable to {expect} that at least a few 
photometrically identified trains are due to a periodic dimming 
caused by spots that lasted for a few rotations.

\subsection{Iterative solution of the 33-spot problem}
In many-parameter problems the only feasible way to derive
a parameter's marginal distribution is by applying the 
MCMC method {\citep{press}}. 
To find a relaxed-looking state, an iterative approach 
was chosen. In one step only the spots contributing to 
the first half of the time series were allowed to 
vary while the parameters characterizing the remaining spots 
were held fixed. In the following step only the spots contributing to
the other half were allowed to vary, with the parameters of the 
remaining spots being those resulting from the step before. 
This process was iterated several times. 
This works because the two sets of spots are virtually decoupled,  
{meaning that} it results in two relaxed solutions that comprise 
all spots. In the end, for a given non-spot parameter, for example,
differential rotation, both solutions 
were combined by computing the mean value from both means. 
The variance of this combined mean follows from the two mean values, 
which are representative for each half of the time series 
and the corresponding variances. This seems to be appropriate as 
long as the parameter space can essentially be reduced to two 
subspaces. 
Nevertheless, even by splitting the 33-spot problem into two 
{(comprising 16 and 17 spots, respectively), it is computationally 
demanding to achieve relaxed MCMC solutions.}
The results presented were obtained by 
running up to 128 Markov chains in parallel over many days.

\subsection{Results of the spot modeling}

With the notable exception of a few singular dips that remain 
unaccounted for by any persisting dark surface feature, 
it is possible to fit each major dip in the light curve by 
the effect of at least one of 33 spots (Fig.~\ref{fig_lc}). Only in {about $25\%$} 
of all cases {does} the spot period exceed 20 days. 
The residuals are at 0.16 mmag, {which means that} {${\sim}87.5\%$} of the 
photometric variance {can be explained by} dark spots. 

The result of the Bayesian parameter estimation 
is the posterior probability distribution across 
the parameter space. By marginalization, {that is, by} 
throwing away the information on correlations 
between parameters, we derived the marginal distributions of
the parameters. Each marginal 
distribution reveals the probability disstribution of a parameter, 
irrespective {of} the values {that} all other  parameters may take on. 
It can be comfortably summarized by its expectation value and 
credibility interval(s). 
The most interesting marginal distributions are 
presented in Figs.~\ref{fig_incl}--\ref{fig_kappa}.

We reall that any Bayesian 
confidence region reflects the elbow room 
of the underlying theoretical model constrained by the data, nothing else{.} 

\subsubsection{Inclination}

\begin{figure}
\resizebox{\hsize}{!}{\includegraphics{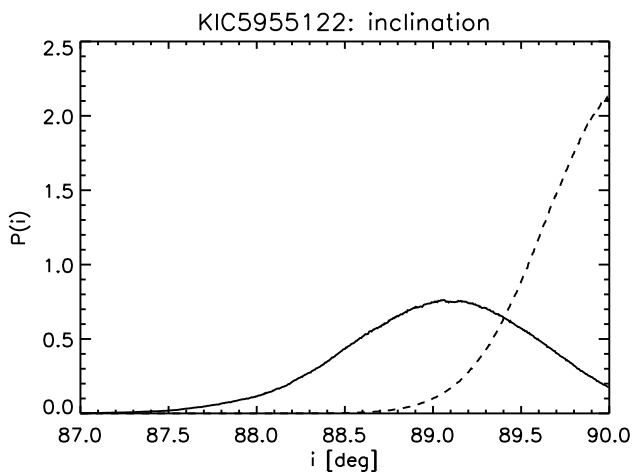}}
\caption{Stellar inclination $i$ from photometry. 
The solid curve is more representative of the first half of the 
light curve, the dashed curve more representative of the second half.}
\label{fig_incl}
\end{figure}

The photometrically estimated inclination, $i$, is surprisingly high, 
nearly $90^\circ$ 
. The two marginal distributions of the inclination, 
computed from the $\cos(i)$ parameter, are shown in Fig.~\ref{fig_incl}. 
The overall inclination, computed from combining the mean values and variances from 
the two halves of the light curve, is {$i = 89.0 \pm 0.5^{\circ}$.} 

A cautionary note is in order. 
The {close-to} 90$^\circ$ {solution for the} inclination might in principle be an artifact of the spot modeling. 
Because of the many spots involved, the hemisphere {of a} 
spot  cannot be estimated by trial-and-error. 
To prevent {spots from overlapping,} the 
sign of a starting spot latitude was simply assumed to alternate with rising 
absolute value of the latitude. The only case where the hemisphere 
of a spot does not matter is in the edge-on (equator) 
view of the star. Perhaps this is the reason for the preference 
of a very high inclination. 

However, the value of {$v \sin i =6.5 \; {\rm km \;s}^{-1}$ obtained from \cite{bruntt12}}    
with an asteroseismic radius
of $R\approx 2 R_\odot$, yields an inclination of nearly 
90$^\circ$, which is perfectly consistent with the value obtained from the spot modeling.

\subsubsection{Differential rotation and highest latitudes}

\begin{figure}
\resizebox{\hsize}{!}{\includegraphics{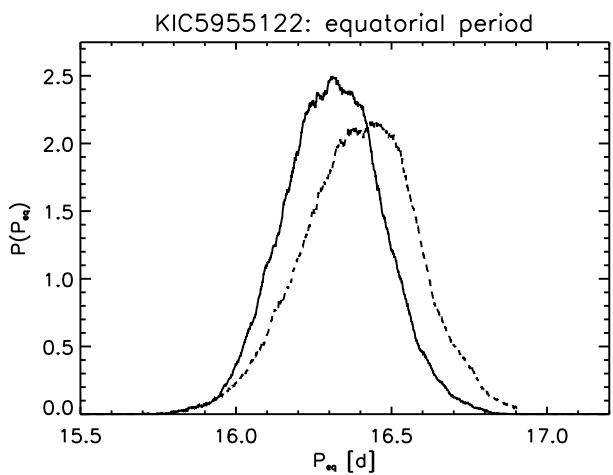}}
\caption{Stellar equatorial rotation period $P_{\rm eq}$. 
The solid (dashed) curve is more representative of the first (second) half of the 
light curve.}
\label{fig_Peqrot}
\end{figure}

\begin{figure}
\resizebox{\hsize}{!}{\includegraphics{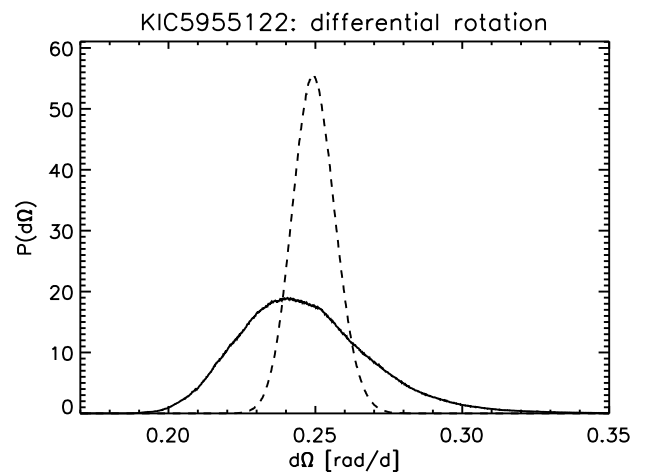}}
\caption{Stellar equator-to-pole differential rotation {$\rm d\Omega$}. 
The solid {(dashed)} curve is more representative of the first {(second)} half of the 
light curve.}
\label{fig_diffrot}
\end{figure}

\begin{figure}
\resizebox{\hsize}{!}{\includegraphics{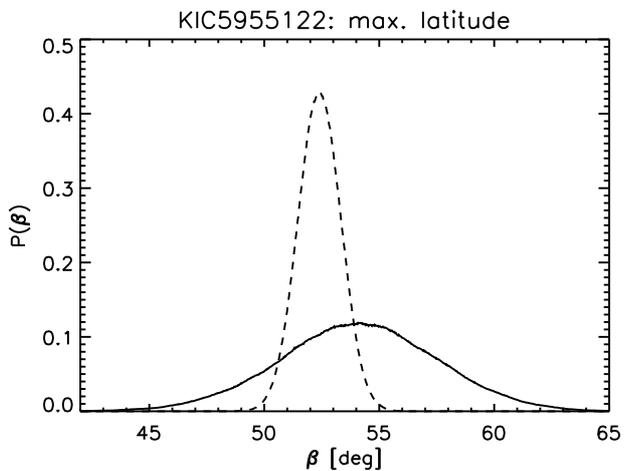}}
\caption{Latitude of the slowest spot. 
The solid {(dashed)} curve is more representative of the first {(second)} half of the 
light curve.}
\label{fig_sin2}
\end{figure}

A lower limit {on the} differential rotation {of} 0.16 rad/d  follows from 
the periods of the fastest (16.4 days) and the slowest spot (27.7 days). 
When we rely upon the photometrically ill-determined
spot latitude, constrained by the period via the parameterized 
$\sin^2$-law of differential rotation, we obtain 
an {\em \textup{extrapolated}\/} equator-to-pole 
differential rotation of $0.25\pm 0.02$ rad/d. Both contributing 
marginal distributions are depicted in Fig.~\ref{fig_diffrot}. 
Because the fastest spot was fixed to the equator, its period measures 
the equatorial rotation period: $16.4 \pm 0.2$ days (Fig.~\ref{fig_Peqrot}). 
By a quirk of fate, this equatorial spot proved to be ephemeral. 
Constituting a train of three consecutive dips and influenced by 
a fast spot area evolution, the spot rotational period is ill-defined. 
{However,} the {second} -fastest spot, with a period of 16.9 days, is represented 
in the light curve {already by} ten consecutive dips.
The slowest spot reaches (in the mean) a latitude of 
{${\sim} 53^\circ$} (Fig.~\ref{fig_sin2}). 
The  corresponding {value for the adopted description of the differential rotation}  reads 
$\sin^2(\beta_{\rm s}) = 0.64 \pm 0.04$, but higher latitudes {(up to about 65 degrees)} are not excluded.
With the exception of periods, all farther reaching 
conclusions depend on dubious model assumptions, {for instance,
on} the circular shape 
of the spots, or on the shape of the law of differential surface rotation, 
and are therefore not as trustworthy as periods.

\subsubsection{Spot intensity}

\begin{figure}
\resizebox{\hsize}{!}{\includegraphics{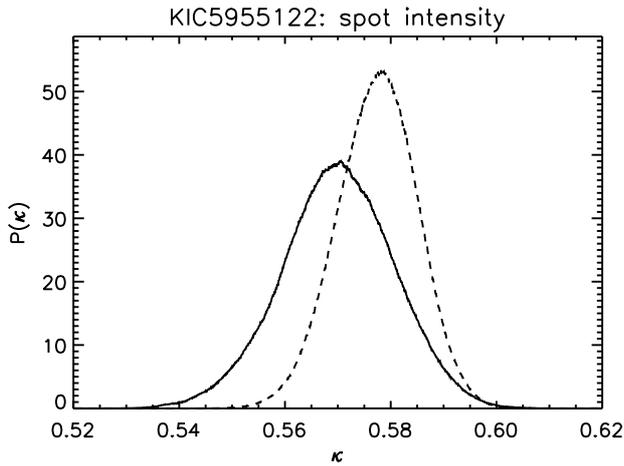}}
\caption{Spot intensity $\kappa$ with respect to the unspotted photosphere. 
The solid {(dashed)} curve is more representative of the first {(second)} half of the 
light curve.}
\label{fig_kappa}
\end{figure}

The marginal distribution of the spot rest intensity, expressed in units of 
the  unperturbed {stellar} surface intensity, {$\kappa$,} 
is depicted in Fig.~\ref{fig_kappa}. {According to this,} the spots are  roughly 
half as bright as the unspotted {stellar} photosphere, {with $\kappa = 0.57 \pm 0.01$.}
Irrespective of the drop in temperature, we applied the same
LD prescription to compute the 
photometric effect of a spot as for the hotter unperturbed surface.

\section{Mean field dynamo model}

It is reasonable to assume that the flux-transport dynamo can operate in other solar-like stars, 
although important differences with the solar case might occur for a strong meridional flow
or a different rotation rate. 

It is convenient to introduce a  Reynolds number of the rotation
as follows{:} $C_\Omega = {R_\ast^2 \Omega_{\rm eq}}/{\eT}$, 
where $R_\ast$ is the stellar radius and $\Omega_{\rm eq}$ {is} the equatorial rotation rate. 
For \kic  it easy to realize  that 
$C_\Omega$ is approximately 1.4  times the solar value. Indeed, 
the turbulent diffusivity{,} $\eT${,} scales as the square of the convective velocities, 
$\Omega_{\rm eq}/\Omega_\odot  \approx 1.5$ from the discussion in the previous section, and $R_\ast \approx 2 R_\odot$.
By similar arguments it is possible to estimate that  the Reynolds number of the flow, $C_u = R_\ast U/\eT${, with $U$ being}
the typical strength of the meridional circulation at the bottom of the convection zone, is about $1.5$ times  the solar value 
{of} $C_u = 400$ \citep{boso}. {Note that because the latitudinal structure of the meridional circulation is still a subject of debate \citep{zao, bedi} we assumed the single-cell
model that is most often used in the literature \citep{bo02, dik09}}.

On the other hand, although the typical dynamo numbers of the rotation and meridional circulations 
are not very different from the solar values, 
the surface differential rotation is about three times stronger than that of the Sun, 
and therefore a numerical approach is essential to extract all the 
relevant information on the topology and dynamics of the toroidal field. This is the main subject of this section.

\begin{figure}[htp]
  \includegraphics[height=7cm ]{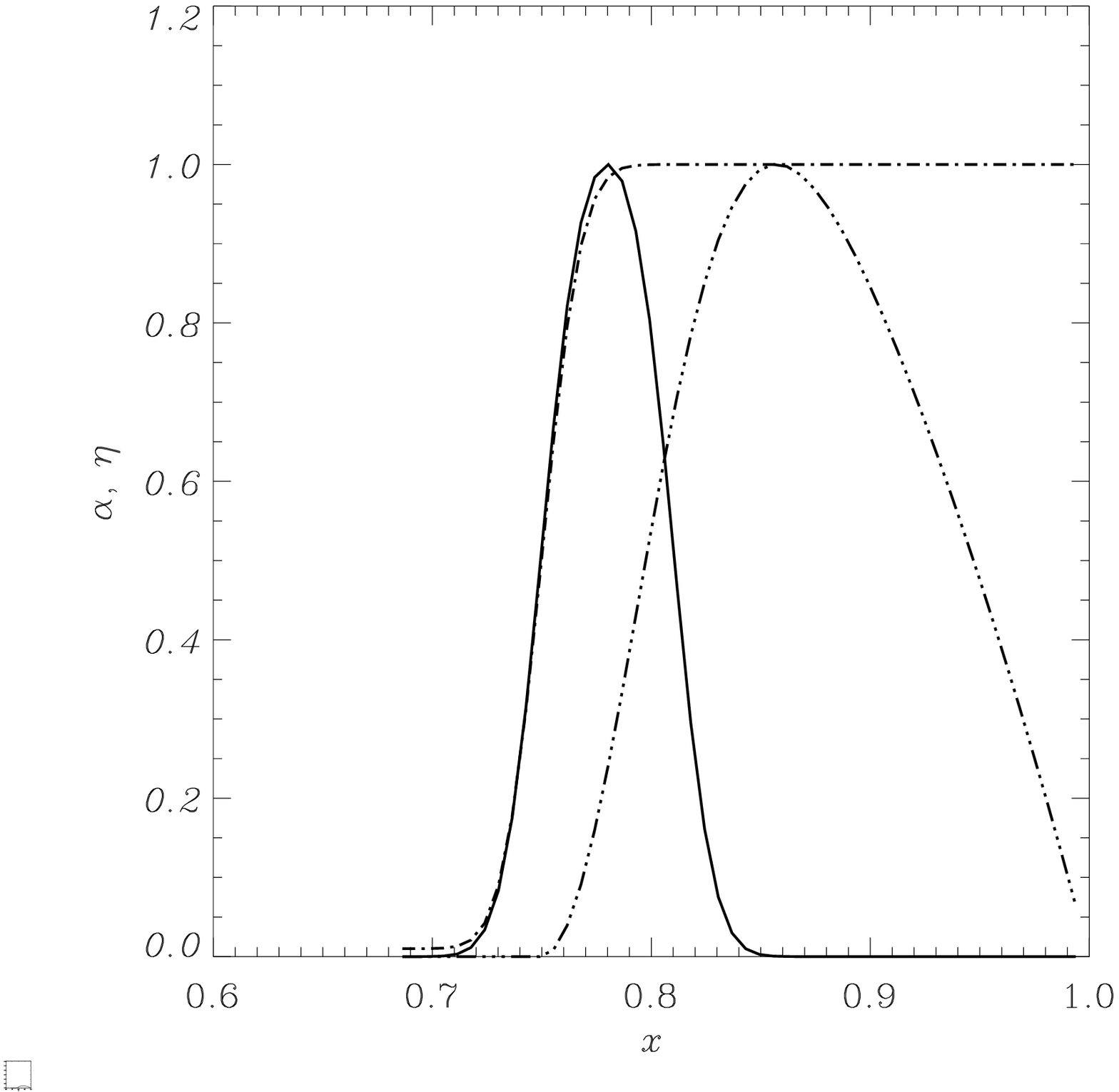}   
  \caption{
  $\alpha$-effect (solid line), turbulent diffusivity (dot-dashed line), and (minus) the function $\psi(x)$ (dot-dot dashed line) used in the computation.
  The maximum of the  $\alpha$-effect corresponds to the location of the convective zone.
    \label{al}}
\end{figure}

\begin{figure}[htp]
  \includegraphics[height=6cm ]{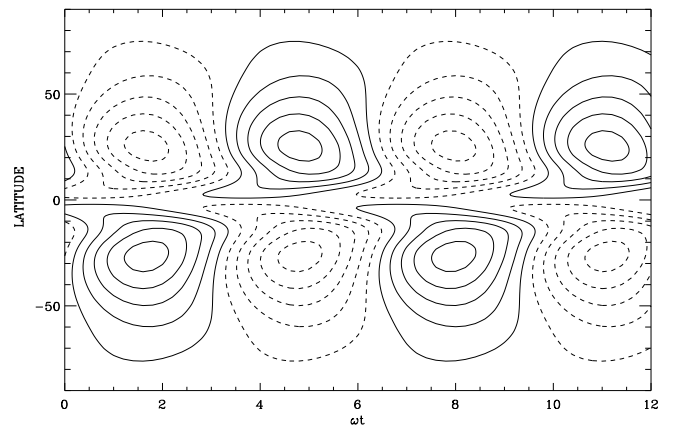}   
  \caption{Butterfly diagram for a reference solution with  
 $C_\alpha = 6.35$, $U = 30 \; {\rm m/s}$ , $\eT = 6.6 \cdot 10^{11} \; {\rm cm^2 s^{-2}}$, and an activity cycle of 21 {years}. 
The toroidal field is evaluated at the bottom of the convection zone. 
 The ratio $B_r/B_\phi = 0.0048$ at the base of the convection zone.} 
\label{bt}
\end{figure}
\begin{figure}[htp]
  \includegraphics[height=6cm ]{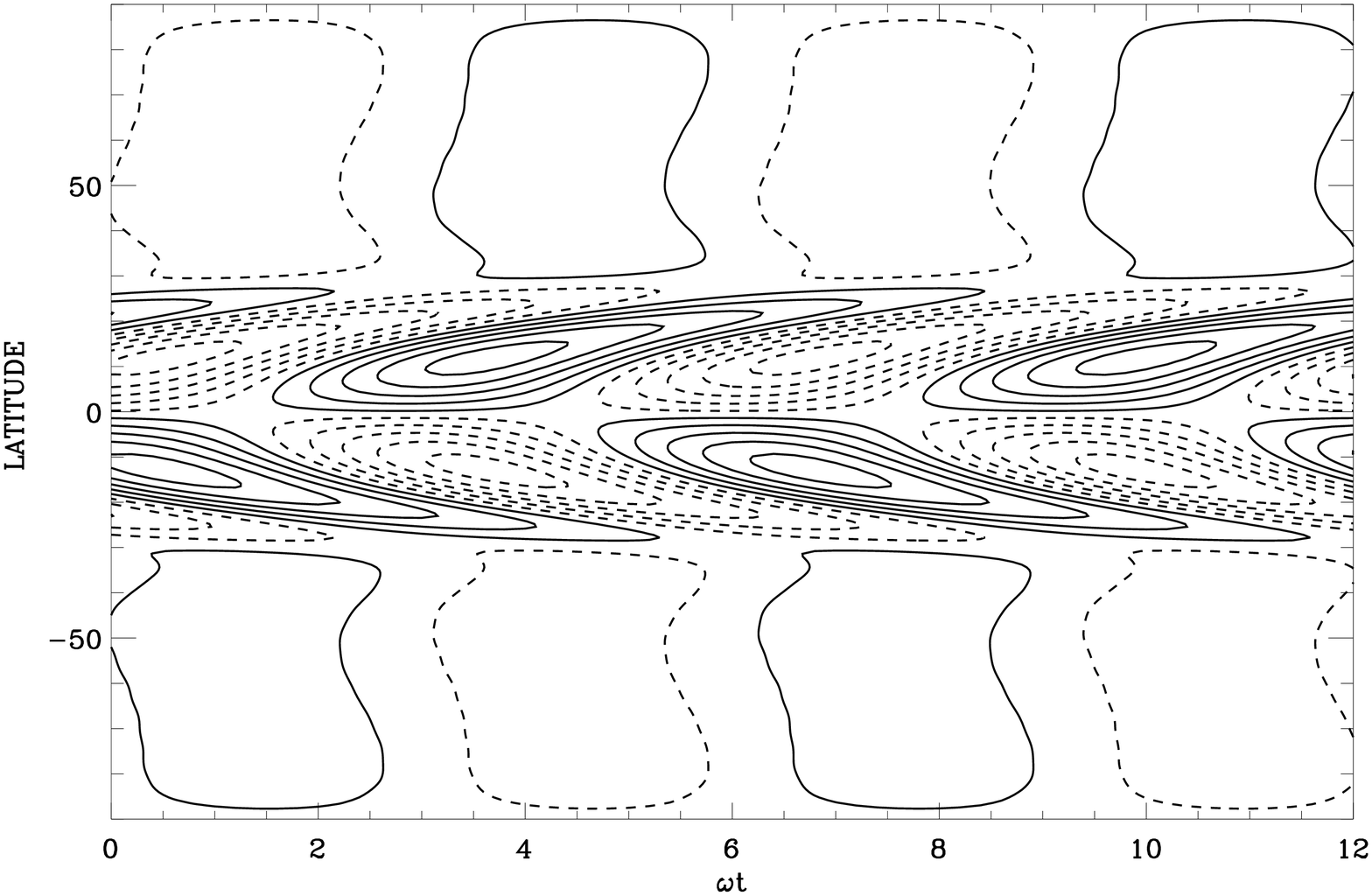}   
  \caption{Same parameters as in Fig.~\ref{bt}, but {with} $C_u=800$. 
In this case, $C_\alpha = 7.33$, while  $U=37  \; {\rm m/s}$. The activity cycle is 17 years. }
 \label{bt2}
\end{figure}

\subsection{Basic equations}\label{basics}
In the following we briefly review the basic ingredients of the advection-dominated dynamo following the discussion in  \cite{boso}.
We write the magnetic induction equation as
\be\label{induction}
{\partial {\bm B} \over \partial t} = {\bm \rot} ({\bm V}
\times {\bm B} + \alpha {\bm B}) - {\bm \nabla} \times \left( \eT {\bm \nabla} \times {\bm B}\right),  
\ee 
where $\eT$ is the turbulent diffusivity.  In spherical symmetry the magnetic field ${\bm B}$ 
and the mean flow field ${\bm V}$ read 
\ba
&&{\bm B} = B_\phi(r,\theta,t)\ephi
+ {\bm \nabla}\times [ A(r,\theta,t) \ephi],\\[2mm]
&&{\bm V} = {\bm u}(r,\theta) + r\sin\theta \Omega(r,\theta)\ephi,
\ea
where $B_\phi(r,\theta,\phi)\ephi$ and ${\bm \nabla} \times [A(r,\theta,t)\ephi]$ 
are the toroidal and poloidal components of the magnetic field.
In this formalism the meridional circulation ${\bm u}(r,\theta)$
and differential rotation $\Omega(r,\theta)$ are the poloidal and toroidal components
of the global velocity flow field ${\bm V}$.
The $\alpha$-effect is always antisymmetric with respect to the equator, so that we write
\be
\alpha = \frac{1}{4}\, \alpha_0 \cos\theta\Big [1+{\rm erf}\Bigl(\frac{x-a_{1}}{d}\Bigr)\Big ]\Big [1-{\rm erf} \Bigl(\frac{x-a_{2}}{d}\Bigr)\Big ],
\label{aa}
\ee
where $\alpha_0>0$ is the amplitude of the $\alpha$-effect, $x=r/R_\odot$ is the fractional radius, 
$a_1=0.75$, $a_2=0.81$ and $d=0.025$ define the location and the thickness of the 
turbulent layer. We assume that below the tachocline the turbulent diffusivity is lower by  few orders of magnitude
than the value attained in the bulk of the convection zone. We can conveniently represent this transition with 
the following functional form: 
\be\label{eta}
\eta = \eta_c+\frac{1}{2}(\eta_t-\eta_c)\Bigl[1+{\rm erf}\Bigl(\frac{x-x_{cz}}{d}\Bigr)\Bigr],
\ee
where 
$\eta_t$ is the eddy diffusivity,  $\eta_c$ the magnetic diffusivity beneath the
convection zone{,} and $d$ represents the width of this transition. In particular, we use the values
$\eta_t/\eta_c=10^2$, $d=0.02$ as for the Sun, with $x_{cz}=0.75$ in our case. 

The components of the meridional circulation can be represented with the help of a stream function 
$\Psi(r,\theta)=-\sin^2\theta\cos\theta\,\psi(r),$ 
so that 
\ba
&&u_r= \frac{1}{r^2\rho\sin\theta} \frac{\partial \Psi}{\partial\theta}=  \frac{1-3\cos^2\theta}{\rho r^2} \; \psi(r),\\
&&u_\theta= -\frac{1}{r\rho\sin\theta}\frac{\partial \Psi}{\partial r}=
\frac{\cos\theta \sin\theta} {\rho r} \; {\dd\psi(r)\over \dd r}
\label{13}
,\ea
with the consequence that the condition $ {\bm \nabla} {\bm \cdot} (\rho {\bm u}) = 0$ is automatically fulfilled. 
In particular, a positive $\psi$ describes a cell that circulates clockwise in the northern hemisphere, meaning that the flow is 
poleward at the bottom of the convection zone and equatorward at the surface.

Following the analysis of the previous section, 
the differential rotation is assumed to be solar-like and
we write
\begin{equation}
\Omega(r,\theta)=\Omega_c+\frac{1}{2}\Bigl[1+\mathrm{erf}
 \Bigl(\frac{x-x_{cz}}{d_c}\Bigr)\Bigr]
\bigl(\Omega_s(\theta)-\Omega_c\bigr), \quad 
\label{eq3}
\end{equation}
\noindent 
where 
$\Omega_c$ is the uniform angular velocity of the radiative core,
$\Omega_s(\theta)=\Omega_{\rm eq}- d\Omega\cos^2  \theta$
is the latitudinal differential rotation at the surface, and $d\Omega$=0.25 rad/d.
To estimate  $\Omega_c/\Omega_{\rm eq}$ 
we used the approach of \cite{garaud} where the coupling between radiative interior and 
convective envelope can be modeled as a spherical magnetized Couette flow.
In our case, this led to $\Omega_c/\Omega_{\rm eq}= (1-d\Omega/3\Omega_{\rm eq})= 0.78$
(note that for the Sun  this approximation provides $\Omega_c/\Omega_{\rm eq}=0.93$ as inferred by helioseismology).

Assuming a standard isotropic mixing-length theory, one can show that to a very good approximation,   \citep{durney,boge} 
the stream function can be described as  
\be\label{dupsi2}
\psi\approx - \frac{5 \rho r }{2\Omega_{\rm eq}} \langle u_r^2\rangle
,\ee
so that it is possible to determine the stagnation point directly from the underlying stellar model {\citep{boge}}.
Note that a negative $\psi$ describes a flow that is poleward at the surface and 
equatorward at the bottom of the convection zone, as we observe in the Sun.
For the actual calculations we employed the following representation for the stream function:
\be
\psi=C \; \left[1-\exp{\left(-\frac{(x-x_b)^2}{\sigma^2}\right)} \right] (x-1) \; x^2 
,\ee
where $C$ is a normalization factor, $x_b=0.75$ defines the  penetration of the flow, $\sigma=0.07$ {and} measures how fast 
$\langle u_r^2\rangle$ decays to zero in the overshoot layer and the location of the stagnation point.
The density profile can be modeled as $\rho=\rho_0 \Bigl(\frac{1}{x}-x_0\Bigr)^m$
with $m=1.2$ and $x_0=0.85$, although our results are not strongly dependent on these values.

\subsection{Results}
We solved Eq.~(\ref{induction}) by means of the code CTDYN described in \cite{jouve10}, which employs a pseudo-spectral decomposition of the induction equation to determine
the critical dynamo number $C_\alpha = \alpha_0 R_\ast/\eT$ in the kinematic regime (see \cite{jouve10} for details). 

By scaling the solar dynamo solution discussed in \cite{boso}, we have that $C_\Omega \approx 6 \cdot 10^4$, $C_u \approx 600$,
and $C_\alpha = 6.35$, $U = 30 \; {\rm m/s}$ with $\eT = 6.6 \cdot 10^{11} \; {\rm cm^2 s^{-2}} $, and an activity 
cycle of about 21 {years is found}.

A typical butterfly diagram for this type of dynamo action is depicted in Fig.~\ref{bt}. It is interesting that there  are two branches, one equatorward, another poleward, starting a lower latitude. 
This is expected because the latitudinal shear is much stronger in this case than for the Sun, and it tends to produce a poleward migration with a positive $\alpha$-effect in the northern hemisphere. 
Moreover, the toroidal field belts are confined to latitudes lower than {${\sim}60^\circ$} 
although the $\alpha$-effect is not suppressed in latitude by the usual $\propto \sin^2\theta$ term. 

We intepret this result as a nontrivial confirmation of the analysis of the previous section, 
according to which spots are distributed mostly below  ${\sim}60^\circ$ in latitude. 
We  argue that this property is a consequence of a flux-transport dynamo action, which confines the active regions at low latitudes. The predicted length of the activity cycle is perfectly consistent with the possibility, discussed in the first section,{} that this star
has undergone a transition from a maximum of the activity to a quiet state during the Q0--Q16 run. 

What happens if we increase the Reynolds number of the flow? In this case, the solution for $C_u=800$ 
is shown in Fig.~\ref{bt2}, where the poleward migration clearly
dominates. 
The very narrow and sharp latitudinal extension of the activity region 
seems not to be strongly supported by our spot model solution, although here polar spots would not be ruled out.

\section{Conclusions}
\kic is an interesting laboratory to test the possibility that a flux-transport dynamo is operating in other solar-like stars.
The strong modulation  observed in the light curve during the 
Q0-Q16 period can be interpreted as a transition from a very active state, maybe a maximum of the activity,  toward  a  quiet state, as supported by a consistent dynamo model.

Although this star is very evolved, it rotates significantly faster than our Sun, and it is conceivable that during its main-sequence life it has hosted a significant dynamo
action. Most probably, its convection zone is still {deep enough} to produce a strong radial shear at its inner boundary.  For this reason \citep{dik01}, the 
significant toroidal field instabilities are expected to produce an $\alpha$-effect that can be {\it positive} in the northern hemisphere.

On the other hand, we note that the star with an $S$-index of 0.14 shows only weak chromospheric activity. 
This could suggest that the dynamos in these evolved F-type stars generally only produce weak chromospheres. 
This idea is partly supported by the $S$-indexes as a function of $B-V$ color index measured by \citet{1995ApJ...438..269B}. 
Here it is clear that KIC 5955122 with a $B-V$ value of 0.54 is located in a sweet spot where stars with similar color generally show 
very low chromospheric activity --  another star with $B-V$ of 0.54 is Procyon, for example. 

Our findings support the idea that the dynamo action in main-sequence and subgiant stars produces photospheric fields that are very different from what is observed in 
young, fast rotating stars that just approach  the zero-age main sequence.

Even though the photospheric field {in \kic} is characterized by a nontrivial topology,  the activity cycle is expected to be of the order (or slightly lower) than the solar cycle. 
This prediction can in fact be falsified  with long-term monitoring of the $S$-index, and we argue that this is an important opportunity to pursue this possibility to test the flux-transport dynamo. 

The  p-mode pulsations and, in particular, the mixed modes can also provide independent probes of the rotation rate, 
inclination angle, and possibly 
upper limits on radial and latitudinal differential rotation. We hope to address these questions in a subsequent paper. 

\section*{Acknowledgments}
We acknowledge the NORDITA dynamo program {\it Differential Rotation and Magnetism across the HR Diagram} for providing a stimulating scientific atmosphere.
Funding for this Discovery mission is provided by NASA's Science Mission 
Directorate. The authors wish to thank the entire {\it Kepler} team, without 
whom these results would not be possible. CK acknowledge support from the Villum
Foundation. 
{Funding for the Stellar Astrophysics Centre is provided by The Danish National Research Foundation (Grant agreement no.: DNRF106). The research is supported by the ASTERISK project (ASTERoseismic Investigations with SONG and Kepler) funded by the European Research Council (Grant agreement no.: 267864).}

\bibliographystyle{aa}

\end{document}